# Semantic Boolean Arabic Information Retrieval

Emad Elabd, Eissa Alshari, and Hatem Abdulkader
Faculty of Computers and Information, Menoufia University, Egypt

**Abstract:** *Arabic language is one of the most widely spoken languages. This language has a complex morphological structure and is considered as one of the most prolific languages in terms of article linguistic. Therefore, Arabic Information Retrieval (AIR) models need specific techniques to deal with this complex morphological structure. This paper aims to develop an integrate AIR frameworks. It lists and analysis the different Information Retrieval (IR) methods and techniques such as query processing, stemming and indexing which are used in AIR systems. We conclude that AIR frameworks have a weakness to deal with semantic in term of indexing, Boolean model, Latent Semantic Analysis (LSA), Latent Semantic Index (LSI) and semantic ranking. Therefore, semantic Boolean IR framework is proposed in this paper. This model is implemented and the precision, recall and run time are measured and compared with the traditional IR model.*

## 1. Introduction

Arabic is one of Semitic languages, spoken by more than 422 million. Information Retrieval (IR) is the process of finding all relevant documents responding to a query from mainly unstructured textual data. The science and practice of storing, searching and founding Arabic information within data is called Arabic IR [6].

The area of IR includes many studies that have been proposed to help users to retrieve information on their interests. The majority of the previously undertaken work describes methods and tools to process English language-based documents. The traditional model for IR framework assumes that each document is represented by a set of keywords, so-called index terms [18]. There are several features that distinguish the Arabic language from other languages. For example, the Arabic language is written from right to left, it has a complex morphological structure, Arabic is polysemous (i.e., the same word may have several meanings) and contains a rich set of vocabulary [19].

Due to the complex morphology, polysemous and the rich set of vocabulary of Arabic language, the traditional IR technologies do not efficiently work with Arabic collections. Therefore, Semantic Web (SW) based IR technologies are nominated to overcome these problems in AIR [10]. SW technologies will enable machines to comprehend semantic documents and data. It can assist the evolution of human knowledge as a whole [6]. It draws conclusions about the Web page and improves the existing Web with machine-interpretable metadata that allows a computer program to understand what a Web page is about. Therefore, IR techniques can be improved using SW technologies.

Ontology is one of the most important knowledge representation techniques in SW. Kumar *et al.* [12] define ontology as "provide semantic for understanding the meaning of data". Ontology is an explicit specification of a representational vocabulary for a shared domain of discourse definitions of classes, relations, functions, constraints and objects [14]. The main purpose of building domain ontology is to mimic how the human brain keeps the semantics stored [2]. The Web Ontology Language (OWL) is a well-known class of ontology [20].

This paper aims to improve Boolean Information Retrieval (BIR) based SW techniques. The proposed model includes the ontology merged with the traditional IR model. It uses ontology to extract Reference Concept (RC) for each term in the collection and in the query. Therefore, semantic Boolean IR achieved high precision in comparison with traditional model.

This paper is organized as follows: The most related works to semantic models and their related techniques are discussed in section 2. In section 3, the proposed model is explained. The experimental results and analysis are shown in section 4. Finally in section 5, the conclusion is provided.

## 2. Related Work

Semantic search approaches are complex because their diversity and large number of dimensions involved in the information search task.

Froud *et al.* [9] use the well-known abstractive model Latent Semantic Analysis (LSA) with a wide variety of distance functions and similarity measures to measure the similarity between Arabic words, such as the Euclidean distance, cosine similarity, jaccard coefficient and the Pearson correlation coefficient. They used LSA with and without stemming in two different dataset to know how stemming impacts the meaning. The results show that when LSA model tries to measure the similarity between two different words with the same root, the use of stemming affects negatively in obtained results.

Fernández *et al*. [8] attempt to bridge the gap between the IR and the SW communities in the understanding and realization of semantic search. They proposed the generation of a novel semantic search model that integrates and exploits highly formalized semantic knowledge in the form of ontology's and KBs within traditional IR ranking models. Table 1 summarized the most known approaches that integrate the SW technologies with IR and their limitations. From the table, there is a big gap between the classic IR approaches and the SW technologies. One of these problems is the lack of Boolean semantic IR model. Therefore, there is a trend to use semantic technologies to develop Boolean semantic IR model. Besides, the listed approaches show a lack of standard evaluation semantic frameworks, semantic ranking and multimedia based ontology.

Table 1. Limitations of semantic search approaches.

| Criterion | Approaches | Limitation | IR | Semantic |
|---|---|---|---|---|
| semantic knowledge representation | Statistical linguistic conceptualization Ontology-based | No exploitation of the full potential of an ontological language, beyond those that could be reduced to conventional classification schemes | ☒ | Partially |
| Scope | WS limited domain repositories desktop search | No scalability to large and heterogeneous repositories of documents | ✓ | ☒ |
| Goal | - | Boolean retrieval models where the information retrieval problem is reduced to a data retrieval task | ✓ | ☒ |
| Query | Keyword query natural language query controlled natural language query structured query based on ontology query languages | Limited usability | ✓ | ☒ |
| Content retrieved | Data retrieval IR | Focus on textual content: no management of different formats (multimedia) | Partially | Partially |
| Content Ranking | No ranking Keyword-based ranking semantic-based ranking | Lack of semantic ranking criterion. The ranking (if provided) relies on keyword-based approaches | ☒ | ☒ |
| Coverage | - | Knowledge incompleteness | Partially | ☒ |
| Evaluation | | Lack of standard evaluation frameworks | ✓ | ☒ |

Where, ✓Exists, ☒Not exists

Abouenour *et al*. [1] propose a semantic Query Expensive (QE) based on Arabic Word Net (AWN). Their work has two types of experiments conducted; the keyword-based evaluation and the structure-based evaluation. It aimed to confirm the preliminary experiments which showed that the accuracy and the Mean Reciprocal Rank (MRR) have been improved and that semantic QE process is adequate to improve the passage retrieval stage of an Arabic Q/A system. The semantic QE approach improves both the accuracy and the MRR. In addition, in the case where it is combined with JIRS, the resulted approach has obtained an accuracy around 19.51% and 7.85 as MRR [1].

Probability of relevant passage improved because they take into account the semantic and the structure of the question. In the other side, the AWN project does not cover totally the standard Arabic version of AWN. It included Word Net only as ontology, because the other Arabic ontology techniques such domain based ontology is difficult measure. In the same trend of IR query processing, Meiyappan *et al*. [16] present an interactive query expansion methodology using Concept Based Directions Finder (CBDF). They apply Explicit Semantic Analysis (ESA) for the user query and identify the relevant terms based on the content and link structure of Wikipedia.

Hoseini [10] uses a Derivational Arabic Ontology (DAO) to model the Arabic language. The knowledge is retrieved when needed for use in computer based applications mainly. The key idea underlying compositional approach is that the meaning of a sentence can be composed from the meaning of its syntactical constituents. In this work, the semantic representation of Arabic syntactical phrase is function of its constituent words and phrases. The automatic ontology constructions use the list of existing Arabic verbs to generate all its derivations and populate the ontology in an easy and straightforward manner. Their work can be used as the perfect Arabic morphology analyzer. This model needs a lot of study and application to assess the efficiency and performance.

Kumar *et al*. [12] make use of the prevailing resources like lexical database Word Net for English language to semantically annotated documents of various domains. They employed some of IR model techniques such as crowing and parsing with ontology technique to achieve the semantic index. The impact of the new index is not studied experimentally. In addition, there are no formal techniques to determine the context of the word and the relation between the contexts.

Al-Rajebah *et al*. [3] present an approach to build ontological models for Arabic language. The ontological model is applied on Arabic Wikipedia to extract the semantic relations of each article used info box and list of categories and relies upon the semantic field theory. The model is evaluated using insufficient measures: Human judges and precision whilst organizes ontology evaluation methods requires two dimensions: ontology quality criteria (accuracy, adaptability, clarity, completeness, computational efficiency, conciseness, consistency and organizational fitness) and ontology aspects (vocabulary, syntax, structure, semantics, representation and context) [21].

Mazari *et al*. [15] present an approach of automatic construction that is using statistical techniques to extract elements of ontology from Arabic texts by reused information extraction techniques for extracting new terms that will denote elements of the ontology (concept, relation). To analyze the texts of the corpus, two statistical methods were used, the "repeated segments" to identify the candidate terms and "co-occurrence" to the updating of ontology. They formed a domain corpus by the recovery of text from articles of journals and books of the domain and also the collection of documents over the Web.

Beseiso *et al*. [7] evaluate the support of some tools such as Protégé, Jena, Sesame and KOAN for Arabic language. Their results shows that the current tools are not sufficient and didn't cover many aspects of AIR models such as indexing, querying, crawling, etc., Thus, there is a critical need of new tools to be developed to support the Arabic language Natural Language Process (NLP) and retrieval semantically.

Aliane *et al*. [4] develop a project to build an ontology centered infrastructure for Arabic resources and applications. It aims at reusing ontology for creating tools and resources for both linguists and NLP researchers. They used Python language for implementing the extraction system. They opted for a statistical approach, namely the method of repeated segments calculation combined with some prior processing of the texts that comprise: Segmentation, light stemming and stop words elimination.

Al-Khalifa *et al*. [2] present a project for building a framework for recognizing and identifying Arabic semantic opposition terms using NLP armed with domain ontology's. Semantic opposition is also based on the concept of semantic fields/domains. They classified the Holy Quran into: Speech recognition, stop words, morphology analysis and ontology engine. The framework requires usefulness evaluation and effectiveness via the judgment of human experts and through comparing it with more traditional approaches (i.e., dictionaries). Sem Q is a framework that takes as an input a Quranic verse (i.e., sentence) and outputs the list of semantically opposed words in the verse along with their degree of opposition.

Aliane [5] presents an ontology based approach for multilingual information retrieval that has been implemented for Arabic, French and English. Their system is based on knowledge representation formalism, namely semantic graphs, which support domain ontology. The domain ontology constitutes the kernel of the system and is used both for indexing and retrieval. The system has been developed and two kinds of interfaces are offered for the expert user who create, manage and update the ontology and for the end user who searches for documents. Their interfaces are trilingual. The user can work with the language of his choice Arabic, French or English. The difficult task for ordinary people who are not familiar with the ontology. However, the expert people in Arabic is insufficient.

## 3. The Proposed Approach

The proposed model is based on both tradional IR processing and SW technologies to build an Arabic Boolean semanticIR framework. The key idea is to build a semantic inverted index to store not only words but also RC reflex the meaning of these words in there phrases context. The reference ontology concept of a word is determined by getting a major concept links between all the words in the phrase. Therefore, it is based on all the terms of the phrase. In other words, all the words in the same phrase have the same reference ontology concept. The proposed model consists of two main parts: Semantic inverted index construction and semantic query processing and retrieval.

### 3.1. Indexing Phase

In this phase, the semantic inverted index of a collection of documents is built. The algorithm of the index creation starts to manipulate each document of the collection by extracting and preprocessing its phrases one after another. The preprocessing operations on the phrase include the removal of the stop words which are listed in the stop words list and the stemming. These preprocessing operations are standard operations in any information retrieval system. The next operation is the reasoning of the ontology using the set of words that are resulted after the phrase preprocessing operation to get a reference concept from the ontology links between these words. Finally, each word of the phrase is stored in the semantic inverted index in the form [word, reference concept, Doc ID] where the Doc ID is a unique identifier for the document that this phrase and this word are belongs to. Algorithm 1 shows the pseudo code for performing this indexing process. The proposed model with an example as shown in Figure 1.

*Algorithm 1: Semantic inverted index-indexing phase (a collection of documents and ontology)*

*#Let CDoc represents the collection of the documents $\{Doc_1,...,Doc_n\}$, Where $Doc_i \in CDoc$ and n is the number of the documents in the collection.*

*#Let $Doc_i$ represents a document that consists of a set of phrases $\{Phr_{i1},...,Phr_{im}\}$ Where $Phr_{ij} \in Doc_i$ and m is the number of phrases in document $Doc_i$.*

*#Let $Phr_{ij}$ represents a phrase that consists of a set of words $\{w_{ij1}, ..., w_{ijl}\}$ Where $w_{ijk} \in Phr_{ij}$ and l is the number of words in phrase $Phr_{ij}$.*

*#Let Ont represents the used ontology and $RC_{ij}$ represents the reference concept for the words of the phrase $Phr_{ij}$.*

*#Let Doc $ID_i$ is the Doc ID of document $Doc_i$.*
*for each $Doc_i \in CDoc$*
*{*
   *for each $Phr_{ij} \in Doc_i$*
   *{*
     *Remove stop list*
     *Stemming each $w_{ijk} \in Phr_{ij}$*
     *Reasoning the ontology Ont by the words $w_{ijk} \in Phr_{ij}$ and get the $RC_{ij}$*
     *for each $w_{ijk} \in Phr_{ij}$*
     *{*
       *Store [$w_{ijk}$, $RC_{ij}$, Do c$ID_i$] in the semantic inverted index*
     *}*
   *}*
*}*
*Return (semantic inverted index)*

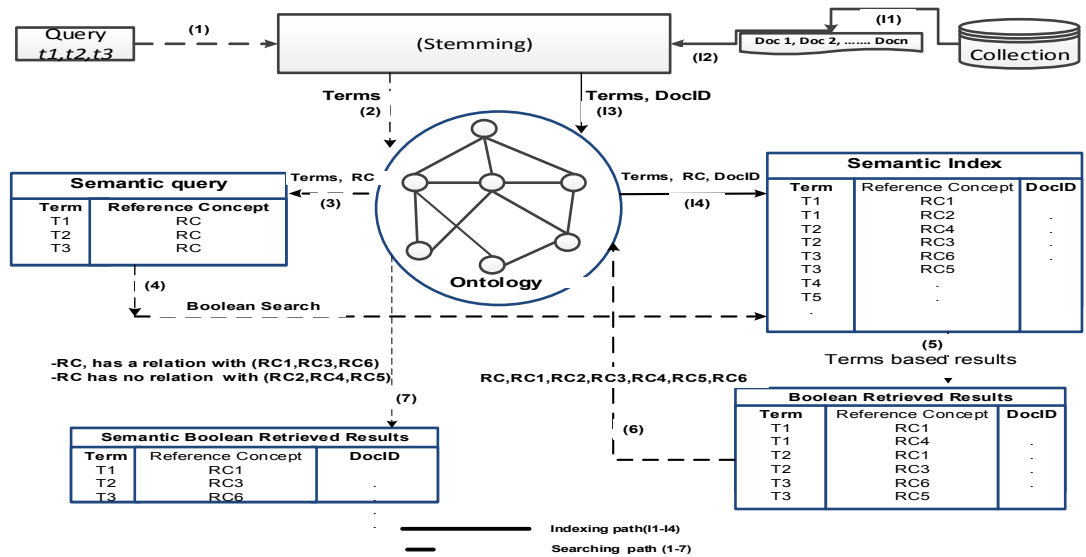

Figure 1. Boolean semantic IR model with informal scenario.

## 3.2. Semantic Query Processing and Retrieval

In this phase, the user's query is processed and the semantic inverted index is used to retrieve the required documents. The query can be a word or a phrase consists of a set of words. In case of only one word, the only preprocessing operation is the stemming and then the information retrieval engine searches in the inverted index for that word and returns to the user the set of documents that contains this word. In this case, if the word is stored in the semantic index with different reference concepts then the returned documents are organized based on the reference concepts to enable the user to select results based on his needs (i.e., in which context he wants his results?). In case of phrase query, the queries is preprocessed by removing the stop words and stemming each word and then check the same ontology which is used in the indexing phase using the set of words of the query phrase and get the reference concept for these words. The previous operation is the same operations that are applied to each phrase on the documents of the collection in the indexing phase. The result of this operation is a set of terms (words) and each term has his reference concept which is the same for all the terms of the query phrase. The next step is to match the terms of the query with the terms of the semantic index. The returned terms will be attached with their RCs. These results are filtered using the ontology by returning only the terms with RCs that have a relation with the RCs of query terms. Finally the filtered results are returned to the user. Algorithm 2 shows the pseudo code for performing this query processing and retrieval process. Figure 1 shows an example of this process where the semantic query reference concept is RC and the equivalent terms have RC1, RC3 and RC6. The filter operation tries to decide if there is a relation between RC and (RC1, RC2, RC3, RC4, RC5, RC6).

*Algorithm 2: Query processing and retrieving (semantic inverted index, ontology, and user phrase).*

*#Let QPh represents a query phrase that consists of a set of words $\{w_1, ..., w_l\}$ Where $w_k \in QPh$ and l is the number of words in query phrase QPh.*
*#LetOnt represents the used ontology and $RC_{ij}$ represents the reference concept for the words of the phrase QPh.*
*#Let $DocID_i$ is the DocID of document $Doc_i$.*
*Read query phrase QPh.*
*Remove stop list.*
*Stemming each $w_k \in QPh$.*
*Reasoning ontology Ont by word$w_k \in QPh$ and get the RC.*
*for each $w_k \in QPh$.*
 *{*
  *Get the $[w_k, RC_i, DocID]$ from the semantic inverted index.*
  *Reasoning the ontology Ont by $RC_i$ and RC.*
  *If there is a relation between $RC_i$ and RC then.*
   *{*
    *Retrieve $[w_k, RC_i, DocID]$ to the user.*
   *}*
 *}*
*Return (List of query words with its corresponding DocID).*

## 4. Experimental Results

The proposed model (semantic Boolean Arabic IR) is implemented using Apache Jena which is a Java based framework for building SW applications [11]. The obtained results are compared with Lucene which is a high-performance, full-featured text search engine library written entirely in Java [13]. The specification of the platform is Intel core2 Duo 2.10 GHz processor and RAM 3 GB on windows 8. We used a sample of Arabic syntactic dataset. For the sake of testing, samples of three different Arabic ontology's are created (علوم-إلكترونيات-طبيعة) using Protégé 3.4.3 software [17]. These ontology's will be used in the creation of the semantic index and the searching process as explained in the proposed technique. The precision of the IR model measures the relevant returned documents from all the returned documents and the recall measure the relevant returned documents from the all relevant documents in the collection. Therefore, the lake of semantic in IR models affects only on the precision but the recall will not be affected. Thus, the precision of the proposed semantic IR model and the traditional IR model is measured using Boolean queries with the three Boolean operators (AND, OR, NOT).

The results in Tables 2, 3 and 4 show the precision of the two IR models by using different queries with OR, AND, NOT operators respectively. In all cases, the precision of semantic IR model is always 100%. This is because the model can detect semantically the required terms and as a result does not return false results. In the other side, the average precisions of the traditional IR model with queries of OR, AND and NOT operators are 43%, 79%, and 44% respectively.

Table 2. Precision of traditional IR and semantic Boolean IR with OR operator queries.

| Queries | Precision | |
|---|---|---|
| | Traditional | Semantic |
| تفاحةOR أبل | 25% | 100% |
| تفاحةOR مانجو | 50% | 100% |
| تفاح OR خوخ | 25% | 100% |
| العــين OR الفراهــيدي | 33% | 100% |
| ألم OR العين | 67% | 100% |
| قناة OR السويس | 50% | 100% |
| قناة OR المستقبل | 50% | 100% |
| Average | 43% | 100% |

Table 3. Precision of traditional IR and semantic Boolean IR with AND operator queries.

| Queries | Precision | |
|---|---|---|
| | Traditional | Semantic |
| تفاحة AND أبل | 50% | 100% |
| تفاحة AND بيضاء | 50% | 100% |
| تفاح AND مانجو | 100% | 100% |
| العين AND للفراهيدي | 100% | 100% |
| ألم AND العين | 50% | 100% |
| قناة AND السويس | 100% | 100% |
| قناة AND المستقبل | 100% | 100% |
| Average | 79% | 100% |

Table 4. Precision of traditional IR and semantic Boolean IR with NOT operator queries.

| Not | Precision | |
|---|---|---|
| | Traditional | Semantic |
| تفاحة أبل بيضاء Not خضراء | 33% | 100% |
| العين Not الفراهيدي | 50% | 100% |
| تفاح Not قناة | 25% | 100% |
| كتاب Not العين | 33% | 100% |
| Average | 44% | 100% |

The high precision of the semantic IR model is costly in terms of time. The semantic index construction time and the search time are increased. This increment is due to the search on ontology to determine the reference ontology concept for each term. Table 5 show the time of traditional IR and semantic Boolean IR with OR operator queries.

The big difference in the time consumed in each case is very clear. Therefore, this problem can be solve by using powerful computers which is already exist and in addition, optimization techniques should be developed to decrease the search time in case of semantic Boolean IR models.

Table 5. Time of traditional IR and semantic Boolean IR with OR operator queries.

| Query | Traditional IR Time (Milliseconds) | Semantic IR Time (Milliseconds) |
|---|---|---|
| تفاحة OR أبل | 3 | 217 |
| تفاحة OR مانجو | 3 | 222 |
| تفاح OR اخضر | 2 | 198 |
| العين OR الفراهيدي | 2 | 137 |
| ألم OR العين | 2 | 282 |
| قناة OR السويس | 2 | 114 |
| قناة OR المستقبل | 2 | 182 |
| Average | 2 | 193 |

## 5. Conclusions

In this paper, a new semantic Boolean Arabic IR model is proposed, this model is based on the use of ontology to represent the relation and the meaning of each word in the index based on it context. The results show that the new approach enhanced the precision and make it 100% in all cases. In contract, the time consumed in the search in the semantic model is very large in compare to the time consumed in the traditional models which is not a big problem nowadays because the existence on powerful computing platform. In the future work, optimization techniques will be developed to decrease the construction time and the search time in the semantic Boolean IR models. In addition, a semantic ranking IR model will be studied and new ranking techniques will be developed.

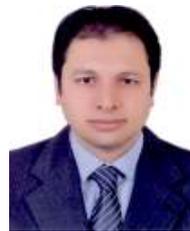

**Emad Elabd** PhD Assistant professor, Department of Information Systems, Menoufia University, Egypt. He got his PhD in the field of Web services compliance over high level specifications at LIRIS, University Lyon1, France, 2011. He received bachelor´s degrees in Electronic Engineering from Menoufia University, Egypt where he did his master's studies in computer science also. His research interests include Web services modeling and analysis with access control and time aspects, Web services (specification, composition), Semantic Web, and Information retrieval.

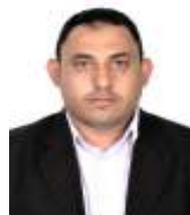

**Eissa Alshari** Master student at Menoufia University, Egypt. He works as a teaching assistant in Center of Computer Sciences and Information Systems Ibb University, Yemen. His main research interest is in information retrieval technologies. He received his BSc in computer science from the national University, Yemen.

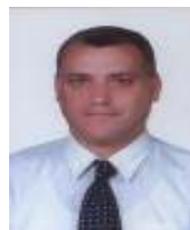

**Hatem AbdullKader** obtained his BS and MSC degrees, both in electrical engineering from the Alexandria University, Faculty of Engineering, Egypt, 1990 and 1995, respectively. He obtained his PhD degree in electrical engineering also from Faculty of Engineering, Alexandria University, Faculty of Engineering, Egypt in 2001. His areas of interest are data security, Web applications and artificial intelligence, and he is specialized in neural networks. He is currently associate professor in the Information Systems Department, Faculty of Computers and Information, Menoufia University, Egypt, since 2004.